# Micro-cracking, microstructure and mechanical properties of Hastelloy-X alloy printed by laser powder bed fusion: as-built, annealed and HIP

Hui Wang[a], Liu Chen[b*], Bogdan Dovgyy[c], Wenyong Xu[b], Aixue Sha[b], Xingwu Li[b], Huiping Tang[d], Yong Liu[a], Hong Wu[a†], Minh-Son Pham[c††]

[a] State Key Laboratory for Powder Metallurgy, Central South University, Changsha 410083, China

[b] Materials Evaluation Centre for Aeronautical and Aeroengine Applications, Beijing Institute of Aeronautical Materials, Beijing 100095, China

[c] Department of Materials, Royal School of Mines, Imperial College London, London SW7 2AZ, UK

[d] State Key Laboratory for Porous Metals Materials, Northwest Institute for Nonferrous Metal Research, Xi'an 710016, China

[*] *Corresponding author. Email contact: leo.chern@yahoo.com*

[†] *Corresponding author. Email contact: wuhong927@126.com*

[††] *Corresponding author. Email contact: son.pham@imperial.ac.uk*

## Abstract

This study analyses literature data to identify optimised print parameters and assesses the consolidation, microstructure and mechanical properties of Hastelloy-X printed by laser powder bed fusion. Effects of post annealing and hot isostatic pressing (HIP) on the microstructure and mechanical properties are also revealed. The susceptibility to the solidification cracking and the as-built microstructure such as precipitation and chemical segregation were predicted by the calculation of thermodynamics phase diagrams. The distribution of solidification cracks throughout the builds was quantified for the as-built, annealed and HIP condition. The assessment reveals variation of cracks towards the bottom, top and free surface of solid builds. This distribution of cracks is found to be associate with thermal gradient and thermal conductivity which were estimated by analytical thermal calculations. While the annealing and HIP both can alter the as-printed microstructure thanks to recovery and recrystallisation, the micro-cracks and pores were only successfully removed by HIP. In addition to the removal, recrystallisation and precipitation in HIP (stronger than in annealing), resulting in optimal mechanical properties including a substantial increase in elongation from 13% to 20%, significant improvement of ultimate tensile stress from 965 MPa to 1045 MPa with moderately high yield stress thanks to precipitation.

## 1. Introduction

Ni-based Hastelloy-X superalloy is widely used in high temperature applications such as aeroengines, for example combustors or fuel nozzles, which generally have complex internal geometries and therefore associated difficulties in fabrication. Additive manufacturing (commonly known as 3D printing) offers many advantages in manufacturing complex components, in particular with reduced lead time, cost and optimisation freedom. Laser powder bed fusion (LPBF) also commonly known as selective laser melting (SLM) is one of the most widely used techniques in additive manufacturing due to its versatility and reliability in fabricating intricate components. In LPBF, the powder is melted selectively according to a sliced 2D profile of a 3D part and then the powder platform is lowered by one layer thickness, repeated layer by layer until the completion of the 3D part [1, 2]. Thanks to its advantages, LPBF has been frequently used to fabricate metallic alloys including Ni-based superalloys [3-8] for various applications in particular for aerospace [9].

Despite the significant potential of LPBF, steep thermal gradient and high cooling rates varying from about $10^4 \sim 10^6 \text{ K} \cdot \text{s}^{-1}$ [10, 11] cause significant concerns regarding the integrity and reliability due to pores, micro-cracks and high internal stresses, limiting the applicability of printed components. Many efforts have been made to improve the quality of AM Hastelloy-X by reducing the porosity. In general, the porosity depends closely on energy density, for example the area energy density [12]. Wang *et al.* investigated the relationship between printing parameters and build quality, demonstrated that a near theoretical density could be achieved when the area energy density of laser beam is above 1.5~2 J mm$^{-2}$ [13]. The energy density dependence of porosity is also reported in other studies of AM Hastelloy-X [12, 14, 15] and other Ni-based superalloys [6, 16].

It is reported that Hastelloy-X alloy is highly susceptible to cracking in LPBF [12]. As compared to the porosity, cracking is much more dependent on the solidification behaviour such as solidification gradient and the freezing range [17] and the thermal gradient [12] which can be estimated by Rosenthal's model [18]. In the early work of Wang *et al.* [13], the density of micro-cracks was found decreasing with the increase of laser power density, manifesting that the higher laser power might be beneficial to lowering the micro-crack density. However, this relationship is not linear and has a limit beyond that the increase in laser power can lead to more cracks because high energy density can lead to more substantial changes in microstructure (such as chemical segregation) in the weld bead and heat-affected zone, reflecting the difficulty in controlling the formation of micro-cracks. To enhance the mechanical performance of AM alloys including the Hastelloy-X, it is necessary to reduce micro-cracks in AM alloys. It was found that micro-cracks can be reduced not only by optimising the print parameters [6, 13] but also via tailoring the alloy composition [12]. For example, Harrison *et al.* [12] demonstrated that the crack susceptibility of Hastelloy-X alloy was reduced by identifying an optimal window of laser power and tailoring the chemical composition. In particular, the tailored chemical composition significantly improved the thermal shock resistance of the alloy [12], highlighting the influential role of microstructure on the crack formation in AM. Moreover, Tomus *et al.* [14] showed that minor chemical elements such as Mn and Si are detrimental to the alloy because of the boundary cohesion due to forming brittle precipitates. This detrimental effect of microstructure was also supported by the work of Marchese *et al.* [19] in which Mo-rich carbides were found along both grain boundaries and crack paths. Therefore, minimising the chemical element segregation towards grain boundaries can lead to the reduction in hot cracking in the alloy [14, 20]. In this study, chemical segregation and microstructure in rapid cooling of Hastelloy-X were predicted by the calculation of thermodynamics phase diagrams (Calphad) that was shown to be capable of estimating the chemical segregation in AM alloys [21]. Solidification parameters (such as the solidification gradient and the freezing range) were also calculated using Calphad to study the susceptibility of the Hastelloy-X to solidification cracking.

In addition to the processing optimisation to improve the quality of AM Hastelloy-X, post-processing such as heat treatment and hot-isostatic pressing (HIP) [9, 22] are also often used to obtain optimal microstructure and eliminate cracks and pores, thereby enhancing the mechanical performance of the alloy for the industrial use of AM. The relief of residual stresses and improvement in consolidation thanks to HIP provide significant beneficial for the mechanical performance of AM components, in particular for superalloys that are often used in demanding environments including high temperatures and high mechanical loads [23]. Therefore, rigorous and quantitative characterisation of micro-cracks and microstructure before and after heat treatment (including HIP) needs to be done to gain in-depth insights and confidence in the application of 3D-printed components made of Hastelloy-X especially in safety-critical application, e.g. hot-section parts of aero-engine.

Although the micro-crack is a great concern in AM Hastelloy-X, it has not been well characterised and investigated quatitatively, in particular throughout the builds, which limits the understanding of cracking mechanisms in the fabrication of parts by additive manufacturing. The printing conditions (such as temperature and thermal conductivity) is varied during LPBF from the bottom to the top and from the interior to free surface of a build [24]. Therefore, a examination of cracks in few locations is not sufficient to characterise the spatial distribution of crack density, limiting the understanding of the relationship between LPBF parameters and crack density for AM Hastelloy-X. X-ray computed tomography (XCT) would be the best technique in quantifying the spatial distribution of defects including cracks. However, XCT is generally expensive and not accessible to wide communities. In addition, XCT has limitation in resolving small defects in particular hair-like micro-cracks found in highly optimised builds [25]. Although microscopic techniques (such as optical and electron microscopy) only provide 2D sectioned views of cracks, they are capable of resolving fine details of micro-cracks. In particular, the characterisation of spatial distribution of cracks via microscopic techniques can be significantly improved by observing many 2D sections throughout the build. Therefore, the detailed optical microscopy imaging at different positions throughout a build, together with automated image processing technique, is viable and reliable to quantify the crack densities inside the build.

In this present study, we carried out detailed quantitative examinations of micro-cracks, microstructure and mechanical behaviour of as-built and heat-treated (including HIPed) Hastelloy-X fabricated via LPBF. We identified a window of optimal printing parameters on the basis of literature review and further developed automated Python codes to quantify the spatial distribution of micro-cracks. Analytical estimates of thermal profile for different pre-heat temperatures and thermal conductivities were done to explain the spatial distribution observed by the crack quantification. The calculation of thermodynamics phase diagrams was also used to understand the mechanisms responsible for cracking in the alloy. In addition, the microstructure and mechanical properties of LPBF Hastellow-X before and after heat-treament (including hot isostatic pressing) were examined to provide in-depth discussions into the cracking mechanism and the influence of post-processing on mechanical properties of the alloy.

## 2. Experimental procedures and micro-crack quantification

### 2.1. Experimental procedures

#### 2.1.1. Powder preparation

The Hastelloy-X alloy powder was supplied by Northwest Institute for Nonferrous Metal Research. Fig. 1 shows a representative micrograph of the as-received powder, and the inset to further details, showing powder particles were spherical with diameters of 15 − 53 μm and a mean diameter of 28 μm. The nominal chemical composition of the as-received powder is shown in Table 1 in comparison with the compositions used in previous studies.

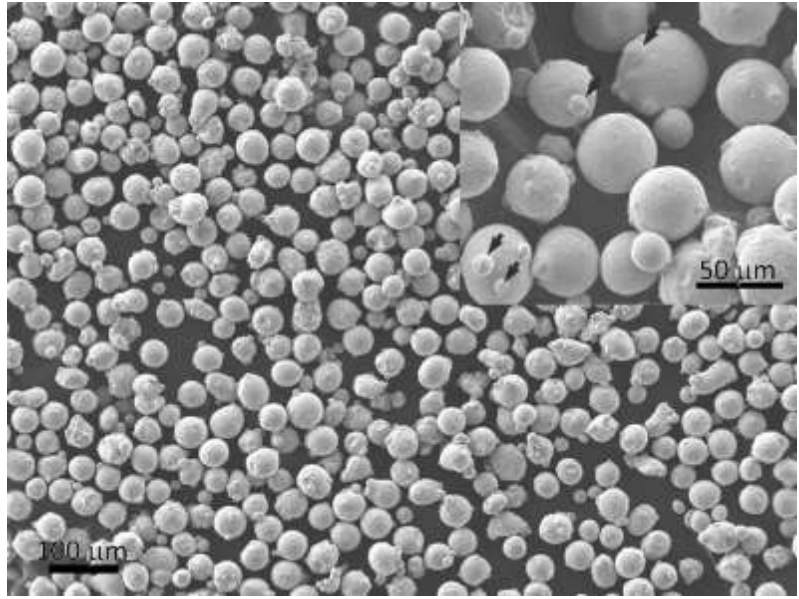

Fig. 1 Secondary electron micrograph showing highly spherical particles of Hastelloy X alloy, with few small satellites (indicated by arrows in the inset).

Table 1 Chemical composition (wt.%) of Hastelloy-X powder used in literature and this study.

| Reference | Ni | Cr | Fe | Mo | Co | W | Mn | Si | C |
|---|---|---|---|---|---|---|---|---|---|
| Wang, *et al*. [13] | Bal. | 20.6 | 18.4 | 8.8 | 1.3 | 0.62 | 0.69 | 0.78 | 0.009 |
| Harrison, *et al*. [12] | Bal. | 21.8 | 18.6 | 9.4 | 1.8 | 1.05 | 0.22 | 0.31 | 0.054 |
| Tomus, *et al*. [18] | Bal. | 21.4 | 18.4 | 8.8 | 1.8 | 0.86 | <0.01 | 0.11 | 0.01 |
| Sanchez-Mata, *et al*.[21] | Bal. | 21.2 | 17.6 | 8.8 | 2.0 | NM | <0.1 | 0.20 | 0.06 |
| This study (by EDS) | Bal. | 23.5 | 20.1 | 7.8 | 1.45 | NM | NM | 1.41 | NM |

*2.1.2. Fabrication*

The fabrication of Hastelloy-X builds was done using an EOS M280 machine equipped with a Yb:YAG fiber laser source of 400 W power capacity, within an Ar protective atmosphere. The powder bed was preheated to 373 K before melting the powder. A meander scan strategy rastered with 16.5° for each layer was employed and the build volume was $10 \times 10 \times 100$ mm$^3$. Previous studies on Hastelloy-X [12, 13, 20, 22] suggest layer thickness of 20 μm and hatch spacing of 90 μm. Laser power and scanning speed were the two variables considered to identify an optimum window that gives minimal porosity and crack density. Fig. 2a and b present a window of the two parameters and measured densities of cracks and porosity reported by Harrison *et al.* [12]. On the basis of the reported data, the Paraeto optimum suggested a laser power of 180 W and a linear energy density of 0.54 J/mm. The optimal scanning speed corresponding to the identified optimal laser power and energy was $0.34 \text{ m} \cdot \text{s}^{-1}$. Consequently, the identified power of 180 W and scanning speed of $0.34 \text{ m} \cdot \text{s}^{-1}$ were used to print Hastelloy-X specimens in this study (Fig. 2b).

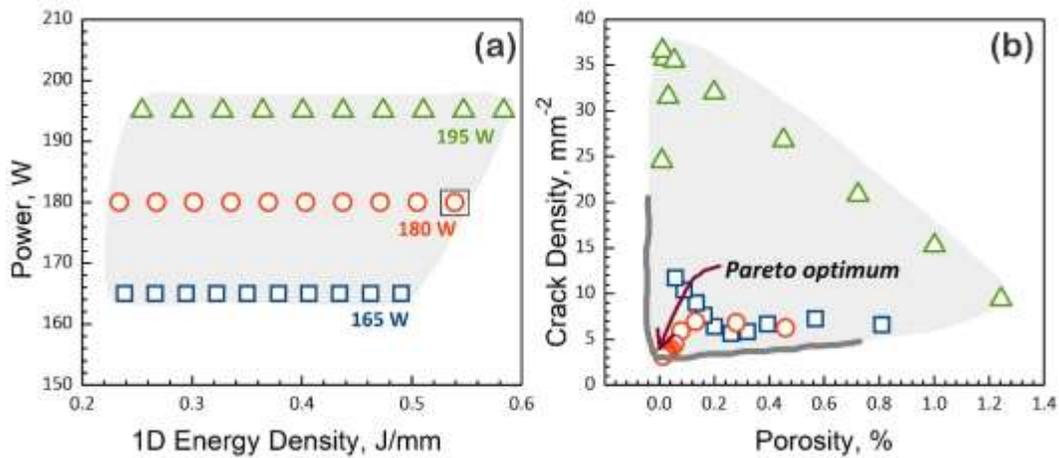

Fig. 2 Identification of optimized power intensity based on Harrison's processing parameters [12]. (a) A window of the laser power and normalized beam energy (note the scan speed was incorporated in the linear energy density). (b) Objective space of crack density and porosity, where the Pareto front corresponds to the circle symbol in (a) (marked by a square).

*2.1.3. Heat treatment and hot isostatic pressing*

Heat treatments by annealing and hot-isostatic pressing (HIP) were carried out to study their effects on the process defects (such as porosity and cracks), microstructure and mechanical properties. The annealing was conducted in vacuum at 1450 K for 4 h, and then furnace-cooled while HIP was conducted at a hydrostatic pressure of 150 MPa, a temperature of 1450 K for 4 h, then slowly cooled in furnace at a cooling rate of about 12 ℃ × min⁻¹.

*2.1.4. Mechanical testing and microstructural characterization*

Fig. 3a shows the sampling layout for mechanical test and microstructural characterization. Optical microscopy (OM) observation position for the quantification of cracks is indicated in Fig. 3b as marked by white crosses. Fig. 3c and d show the geometries of tensile specimens and as-built cuboids.

Vicker's micro-hardness indentation was operated using FM100 with a load of 20 g and holding time of 15 s for each test. The indentation was repeated for more than 15 times for each specimen. Locations of micro-hardness tests for each polished surface are shown in Fig. 3b.

Quasi-static mechanical tensile tests were done using a MTS Landmark test machine at room temperature with a nominal strain rate of $5 \times 10^{-4}$ s⁻¹. Dog-bone specimens (Fig. 3c) had gauge dimensions of $12 \times 2.5 \times 1$ mm³. The surface of specimens was polished and at least 3 tests were done for each material condition to check the repeatability of testing.

Optical microscopic observation was done using a Leica DM4000M microscope. OM samples were prepared by grinding with silicon carbide papers from 600 to 5000 grits, followed by mechanical polishing with Struers OP-S suspension containing 0.06 μm particles. Polished samples were also examined by electron microscopy including electron back-scattered diffraction (EBSD), secondary electron imaging (SE), back-scattered imaging (BSE), and energy dispersive spectroscopy (EDS) using a field emission JSM 7100F scanning electron microscope (SEM) with a voltage of 20 kV. EBSD scanning was operated with a step size of about 1 μm and map size of 500×500 μm², and the data analysis was performed using OIM Analysis (EDAX-TSL). Chemical etching for 5~7 seconds was done for SE scanning using a solution of 36% hydrochloric, 38% acetic and 65% nitric acid in a volumetric ratio of 3:2:2. The aspect ratio of grains (which were defined by a misorientation angle of 15°) was determined as the ratio of the major axis to the minor axis.

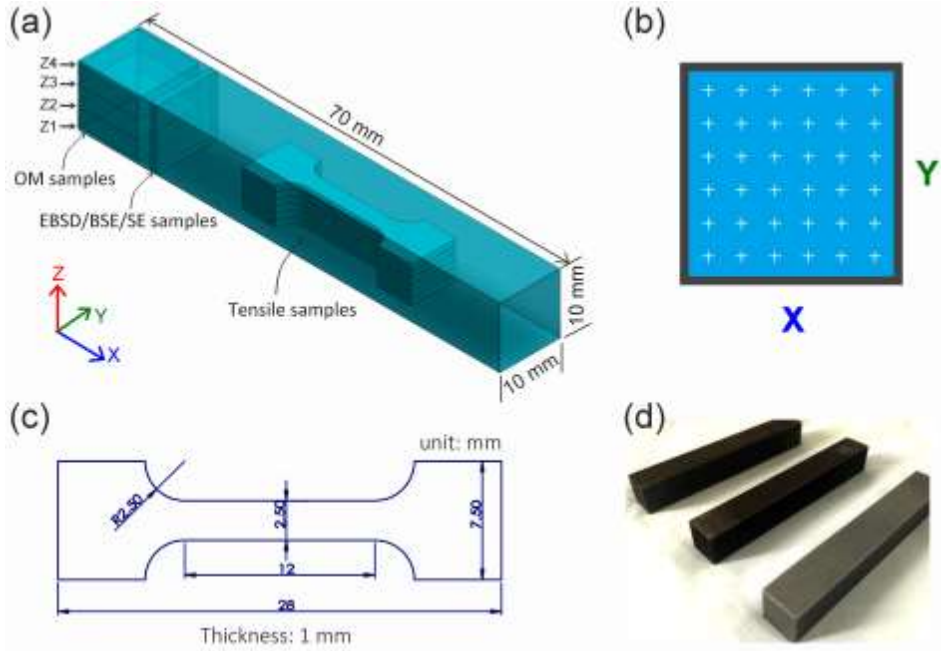

Fig. 3 (a) Schematic illustration of where samples were taken from the printed cuboid. (b) White crosses locating where the OM micrographs taking from at each slice with different height. (c) The tensile sample geometries, and (d) printed Hastelloy-X cuboids.

## *2.2. Micro-crack quantification*

Crack quantification is done on the basis of number of cracks found in OM micrographs. This study used a Python-based package originally developed by Griffiths *et al.* [26] to automatically identify and analyze cracks and the associated statistics from a large number of OM micrographs taken on different locations in a LPBF build to quantify the spatial distribution of cracks. A dimensionless density of cracks is introduced using the Walsh 2D crack density [26, 27], $\gamma$:

$$\gamma = N_A c^2, \tag{1}$$

where $N_A$ is the number of cracks per unit area, and $c$ is the half of the mean of crack length.

The number density $N_A$ could be measured either by direct counting or by grid method, wherein the latter, two sets of perpendicular lines were generated automated to form the grid. The number of intersections between cracks and grid lines per unit length in the two orthogonal directions *I* and *II* is counted and labeled as $P_I$ and $P_{II}$. The crack length per unit area $L_A$ is linked to the number density $N_A$ and the mean crack length $2c$ through $L_A \approx 2cN_A$ [26]. Based on the work of Underwood [28], the crack length per unit area $L_A$ can be equated to the numbers of intersections: $L_A = \frac{\pi}{4}(P_I + P_{II})$. Therefore, $N_A = \frac{\pi}{8c}(P_I + P_{II})$, thus the crack density is given by:

$$\gamma = \frac{\pi c}{8}(P_I + P_{II}). \tag{2}$$

An example of image processing to quantify the densities of cracks is shown in Fig. 4. The orginal OM image of an as-built specimen was converted to grayscale using ($0.17 \times$ Reds $+ 0.28 \times$ Greens $+ 0.55 \times$ Blues) to substantially increase the contrast (Fig. 4a). Some micro-cracks (marked by solid arrows) and minute pores (marked by open arrows) are shown in Fig. 4a. To differentiate micro-cracks which has high aspect ratio feature from porosity, median filtering was operated using a window size serveral times greater than typical crack length. Subsequently, the watershed algorithm was employed to do segmentation based on the difference between the original and filtered images (Fig. 4b). To avoid a mis-identification of cracks from pores or polished stain, a feature was identified to be a crack if the diagonal length of a square that enclosed the feature of interest was equal or longer than a length threshold (which was a length of cracks typically observed by OM, and was found to be 10 μm). The crack shown in the inset of Fig. 4b was thick and contained branches. Therefore, it was subsequently thinned (Fig. 4c) and pruned (Fig. 4d) to ease the measurement of the length of a crack. The pruning process might remove the crack tips which were important to determine the crack length. The removed crack tips were, therefore, recovered afterwards (arrow in the inset of Fig. 4e). Further, the joint of multiple cracks was identified and such cracks were seperated at their joint junction (circle in the inset of Fig. 4e). Finally, the quantification of micro-cracks was conducted by either measuring the length of individual crack or 2D grid method (Fig. 4f).

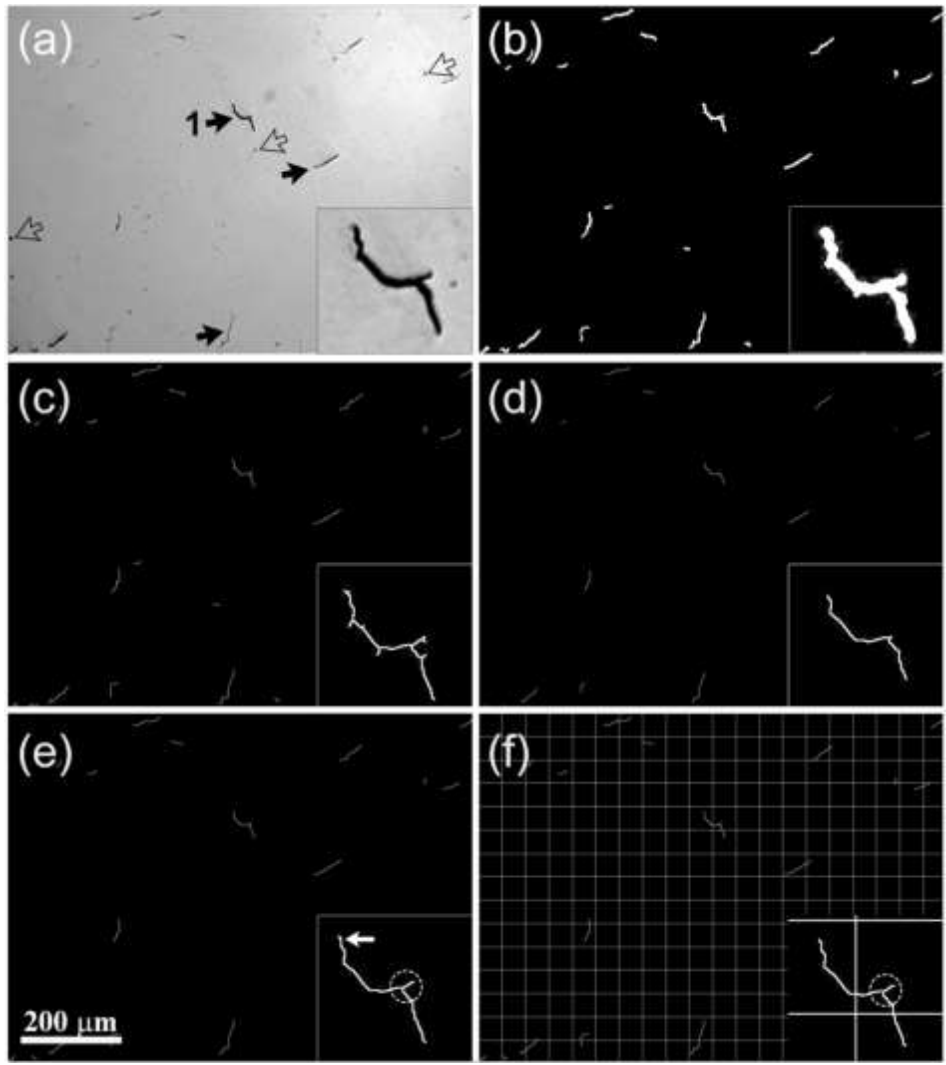

Fig. 4 Image processing to quantify micro-cracks. (a) Gray scale view of original image taken from an as-built specimen (the solid and open arrows highlighting cracks and minute pores, respectively. The inset shows a zoom-in of a crack marked as '1' in (a). (b) Segmented image of (a) by median filtering. (c) Thinned cracks showing some small-crack branches remained. (d) Pruned cracks after trimming the small-crack branches. (e) Final cracks after restoring the tips (marked by the arrow in the inset) of the pruned crack 1 and the decomposition (located by the dotted circle in the inset) of the crack 1 into major and minor cracks to ensure only two end-points in each crack. (f) Cracks in (e) overlapped with grid to count the intersection points per length by Underwood's method [29].

## 2.3 Temperature profile produced by moving heat source

Thermal profile parameters such thermal gradient are influential to the formation of porosity and cracks. Carslaw *et al*. had developed [29] the analytical solution of temperature distribution with a specific heat flux in a semi-infinite solid. With the heat supply from a laser beam over the duration time of $\tau$, the temperature in the solid at the depth $z$ beneath the heat source and time $t$ is given by the following formula assuming that the boundary ($z = 0$) is thermally insulated [29]:

$$T(z,t) = T_0 + \frac{2H}{k}\sqrt{\alpha t}\; \boldsymbol{ierfc}(\frac{z}{2\sqrt{\alpha t}}), \quad z > 0, 0 < t \leq \tau, \tag{3}$$

where $T_0$ is the initial temperature of the powder bed,

$H$ the laser power density $\left(H = \frac{Laser\ power}{Laser\ beam\ area}\right)$,

$k$ the thermal conductivity of the material (given in Table 2),

$\alpha$ the thermal diffusivity of the material $\left(\alpha = \frac{Thermal\ conductivity}{Density \times Specific\ heat\ capacity}\right)$,

$\tau$ is the laser exposure duration time which is calculated as $\tau = \frac{Laser\ beam\ diameter}{Scan\ speed}$,

$\boldsymbol{ierfc}$ is the integral of the complementary error function.

When the heat supply from the laser beam ceased after the duration $\tau$, the temperature is given by:

$$T(z,t) = T_0 + \frac{2H}{k}\sqrt{\alpha}\left[\sqrt{t}\; \boldsymbol{ierfc}\left(\frac{z}{2\sqrt{\alpha t}}\right) - \sqrt{t-\tau}\; \boldsymbol{iefrc}(\frac{z}{2\sqrt{\alpha(t-\tau)}})\right]. \tag{4}$$

with $t > \tau$ and $z > 0$

Values of thermal conductivity and specific heat capacity at calculated temperatures used in calculation are listed in Table 2. The density of $8.22 \times 10^3$ kg/m$^3$ at room temperature was used.

Table 2 Thermal conductivity and specific heat capacity of Hastelloy-X [30].

| Temperature (K) | Thermal conductivity W/(m·K) | Specific heat capacity J/(kg·K) |
|---|---|---|
| 298 | 9.2 | 486 |
| 373 | 11.2 | 487 |
| 473 | 14.1 | 484 |

## 2.4. Solidification cracking prediction and thermodynamic phase diagram calculation

Scheil-Gulliver equation was used to predict the solidification range and solidification gradient that are the key parameters controling the solidification cracking [17]. Solidification range is the temperature range bounded by the liquidus temperature (*i.e.* start of solidification) and solidus temperature (*i.e.* end of solidification) of an alloy. Over this solidification range, the alloy is in a semi-liquid state, *i.e.* mushy zone. During cooling, the mushy zone is subjected to solidification shrinkage and has a potential risk of developing a solidification crack. Another important metric associated with solidification cracking is the solidification gradient, defined by Kou as $d(T)/d(\text{solid fraction})^{0.5}$ at the terminal stage of solidification (typically taken as 0.87 < solid fraction < 0.94) [17]. In his work, Kou demonstrated that if the amount of solidification shrinkage exceeds the lateral growth of solidification cell, liquid film can break down and a crack may form. Both the solidification shrinkage and lateral growth of solidification cells are related to the solidification gradient. A lower absolute solidification gradient results in a higher resistance to solidification cracking.

Calculation of phase diagrams (Calphad) method was performed to evaluate the phase stability at different tempratures and during the temperature changes (solidification and heat treatments). Commecrial software package, ThermoCalc, was used alongside with thermodynamic superalloys database TCNI8 and mobility database MOBNI5, both provided by ThermoCalc. Alloy composition measured by EDS was used for the Calphad calculations. Scheil approximation was used to evaluate the segregation of chemical elements and phase constitution during rapid solidification with a calculation step of 1K. Simulations were carried out for alloy compositions as measured by EDS and reported by Harrisol *et.al* [12]. Both Scheil and equilibrium conditions were used to construct the property diagram of phase constitution from liquidus to room temperatures for the current alloy. This information was used to set up the simulation of the heat treatment. Heat treatment simulation was carried out in Thermocalc Dictra and Prisma modules. Triple phase model of FCC, Sigma and $M_{23}C_6$ phases were used as an input with elemental segregation profile obtained from the Scheil calculation. The modelling domain was set up as a cylinder with a diameter of 600 nm (equivalent to the size of solidification cells). To reduce the computation cost the simulation was carried out in 2 stages. First stage analysed the effect of the heat treatment at 1450 K for 3600 s in ThermoCalc Dictra. Second stage was carried out in ThermoCalc Prisma, a precipitate formation module. Elemental segregation from the 1$^{st}$ stage were used to simulate the cooling of the material from 1450 K to 298 K. EBSD results were used to calculate the grain size and aspect ratio for each scenario. Precipitation was set to be able to form both in the bulk and on grain boundaries, as observed in SEM imaging. As a simplification, precipitates were set to be spherical, while the matrix was set to be isotropic with shear modulus of 77.5 GPa and Poisson's ratio of 0.32 [31], to account for the transformation strain. Dislocation density for both materials was set to $10^{12}$ m$^{-2}$, as a reflection of annealed state [32]. For HIP treatment the cooling rate was set to 0.2 K/s, while for air cooling it was estimated to be 8 K/s [33].

## 3. Results

### *3.1 Thermal profile estimate*

Thermal profile are influential in the microstructure [34] and crack formation during solidification in additive manufacturing. In powder bed fusion, the heat transfer is mainly done via conduction through the solid substrate and surrounding powder. However, the heat condutivity via powder is low due to gaps between powder particles. Therefore, most of heat should be conducted away via the solid substrate and the heat extraction decreases with increasing the distance from the substrate, resulting in higher cooling rates in the bottom part of a solid build than that towards the top of the build. Similarly, the cooling rate near the free surface of a solid build is lower than towards the centre of the solid build due to the presence of surrounding powder particles. In addition to the variation in cooling rates, residual heat generated in previous layers also affects the thermal profile towards the top of a build [24]. The effect of residual heat left in previous layers on the thermal profile of a new melt pool can be studied by using a different initial temperature of the substrate.

To estimate the change in thermal profile with varying the substrate temperature, three different initial temperatures (298 K, 373 K and 473 K) were used, by which to reflect the cold bed, pre-heated bed to 373 K, and bed temperature rise (to 473 K and even higer in practice) due to laser heating with proceeding of printing, respectively. Physical parameters in Table 2 were used for calculation. Eqn. 3 and 4 give the temperature and thermal gradient along the depth at the centre of a heated site at given times corresponding to three initial temperatures (Fig. 5a – c and e – g). The magnitude of thermal gradient was estimated to be very high (about $5 \times 10^6$ K·m$^{-1}$) consistent with other studies which predicted and measured the thermal gradient to be in the order of $10^5 - 10^7$ K·m$^{-1}$ in powder bed fusion [34, 35]. With the temperature increase of initial temperature from 298 K to 473 K, the magnitude of thermal gradient at 300 μm depth and 0.02 s was reduced by about $2 \times 10^6$ K·m$^{-1}$.

To estimate the influence of lower thermal conductivity to temperature distribution when print at the wall area of a build due to poor thermal conduction of adjacent powder, Fig. 5d shows the temperature profile at the same substrate temperature (473 K) but with a lower thermal conductivity as compared with Fig. 5c. The corresponding thermal gradient of the lower thermal conductivity is shown in Fig. 5h. At a given time, the calculation predicts that lower thermal conductivity results in a higher magnitude of thermal gradient as predicted by the Fourier's law of heat conduction [29] with assumption of steady thermal conduction without convection and radiation.

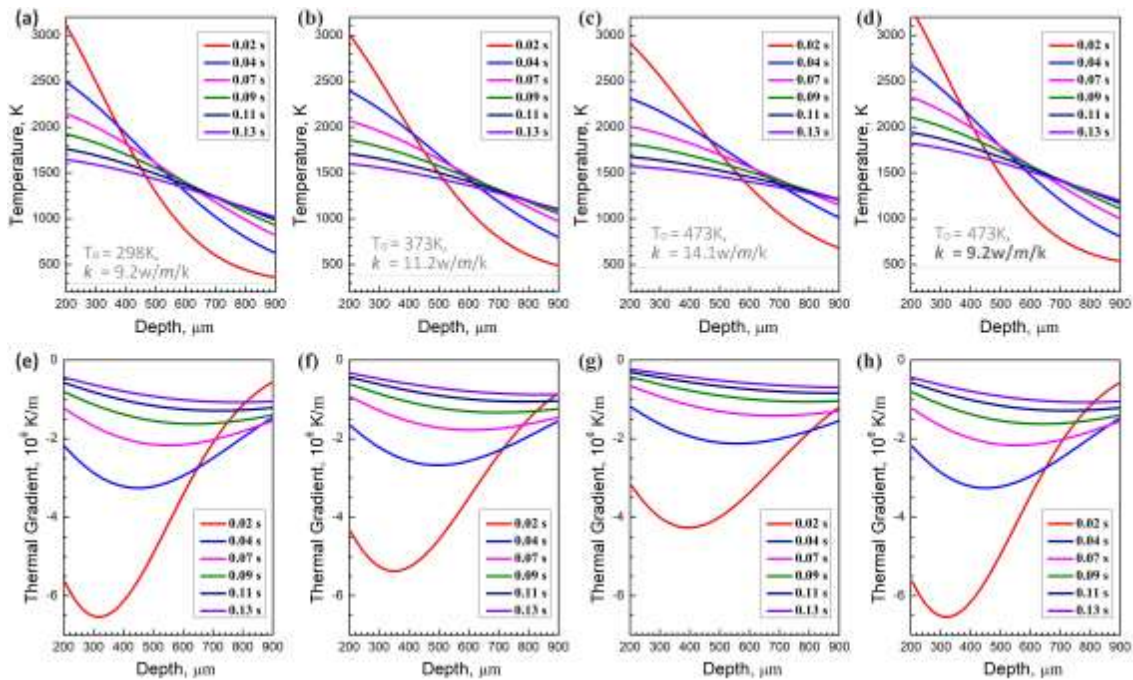

Fig. 5 Influence of different base temperature and thermal conductivity on temperature profile. (a – d) Temperature and (e – h) thermal gradient profiles corresponding to three different base temperature: 298 K, 373K and 473K. The thermal conductivity in (d) is set lower to reflect the case of printing in wall area adjacent to power bed. (Legend refers to the time)

*3.2. Mechanical properties*

Fig. 6a shows the representative engineering stress-strain curves of the material in as-built, annealed (HTed) and HIPed conditions. The as-built specimen showed the yield stress (at 0.2% strain offset) of about 780 MPa and a uniform elongation of about 13%. After annealing, the yield stress dropped substantially (by about 160 MPa), but the uniform elongation does not increase. HIP resulted in excellent combination between strength and ductility: a high yield stress (of about 530 MPa), the uniform elongation of 20 % (which is a significant increases from 13%) and an substantial increase in the ultimate tensile stress to be about 1000 MPa which was higher than that of the as-built condition.

The micro-hardness of the different material conditions along building direction is presented in Fig. 6b. The micro-hardness data were consistent with the tensile strength. Comparing with the as-built one, the annealed condition showed a drastic decrease in hardness while HIPing exhibited hardness comparable to the as-built condition. In addition, the measurement of micro-hardness from the bottom of the build toward the top appears to show a small increase for all three conditions (i.e. as-built, annealed and HIP) but it was within the error ranges. The mechanical properties in present study as well as some reports in literature were detailed in Table 3.

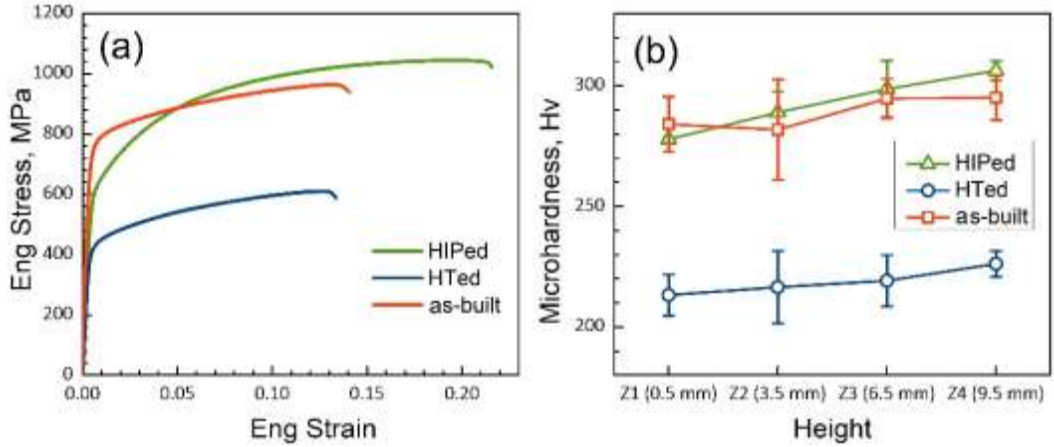

Fig. 6 (a) Engineering stress−strain curves of Hastelloy-X alloy after annealing (HTed), HIP (HIPed curve) in comparision to the as-built condition. (b) Microhardness values measured at different cross sections along the build direction.

Table 3: Summarization of mechanical properties of Hastelloy-X printed by LPBF.

| State | Elastic modulus (GPa) | Yield stress (MPa) | UTS (MPa) | Elongation (%) | Hardness (Vickers) | Ref. |
|---|---|---|---|---|---|---|
| As-printed | 149 ± 9 | 480 ± 10 | 620 ± 15 | 40 ± 1 | NM | [15] |
| HIP | 150 ± 5 | 350 ± 6 | 560 ± 9 | 41 ± 1 | | |
| As-printed | 175 ± 10 | 720 ± 2 | 882 ± 5 | 23 ± 2 | 277.1 ± 3.9 | [12] |
| As-printed | NM | 650 ± 25 | 695 ± 2 | 9 ± 3 | 250 ± 10 | [22] |
| HT | | 410 ± 10 | 666 ± 2 | 23 ± 1 | 212 ± 6 | |
| HIP | | 475 ± 50 | 808 ± 2 | 40 ± 1 | 205 ± 2 | |
| HIP+HT | | 435 ± 25 | 777 ± 2 | 49 ± 1 | 220 ± 5 | |
| As-printed | NM | 814 ± 2 | 930 ± 4 | 35 ± 1 | NM | [9] |
| HIP | | 556 ± 1 | 840 ± 1 | 30 ± 0.2 | | |
| As-printed | NM | 785 ± 18 | 970 ± 14 | 14 ± 2 | 290 ± 5 | This Study |
| HT | | 735 ± 15 | 615 ± 12 | 13 ± 1.5 | 220 ± 5 | |
| HIP | | 635 ± 8 | 1050 ± 10 | 21.5 ± 2 | 290 ± 5 | |

### *3.3. Process defect characterisation*

The material was fabricated with an excellent consolidation with some minute defects (Figs 7 and 8) suggesting the effectiveness of the identification values of the parameters. Most of process defects were identified by electron microscopy to be micro-cracks (Fig. 7), indicating the identified print parameters (Figure 2) were successfully in achieving a high consolidation with minimal porosity. The location of cracks were at the boundaries between different domains of solidification cellular microstructure (as marked by arrows in Fig. 7), suggesting cracks were due to the solidification cracking or chemical segregation-induced cracking during cooling.

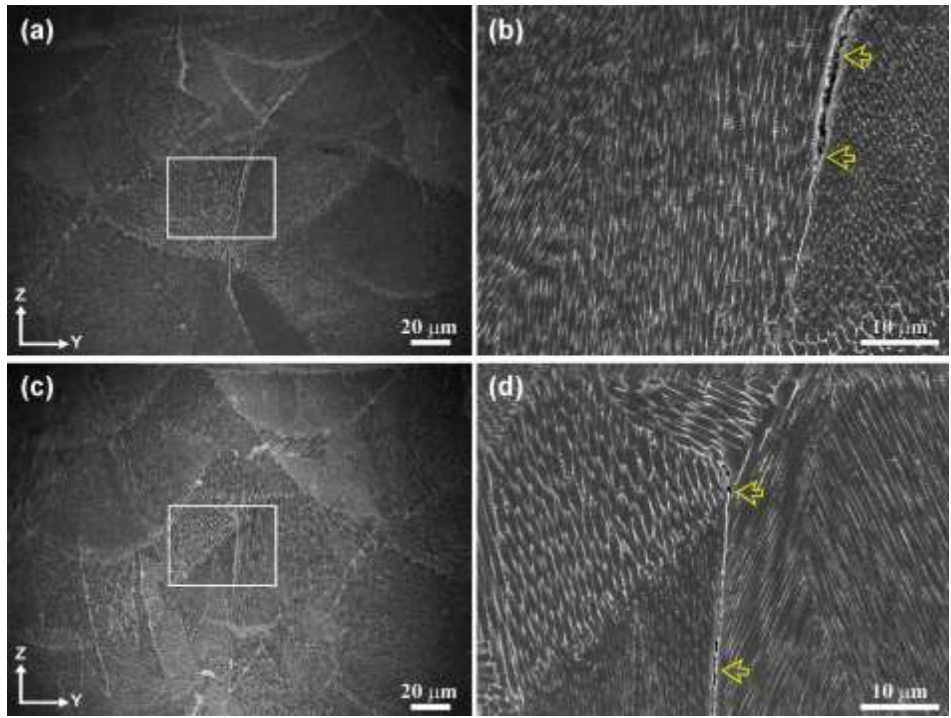

Fig. 7 Secondary electron micrographs showing solidification cracks in as-built specimens. (a, c) Fish-scale microstructure. (b) and (d) Magnified images of the rectangles marked in (a) and (c), respectively, showing solidification cracks at boundaries between solidification cellular domains.

Detailed quantification of the density distribution of micro-cracks is shown in Fig. 8. The as-built specimen showed drastic variations in the crack density along the build direction with higher concentrations near the bottom and top layers. In addition, the density of cracks is much higher near the free surface compared with that towards the centre of the build (Fig. 8a). Almost similar distribution of cracks was seen in annealed condition (Fig. 8b). While there were a lot of cracks density in the annealed condition, the HIP treatment almost removed cracks in all inspected layers (Fig. 8c).

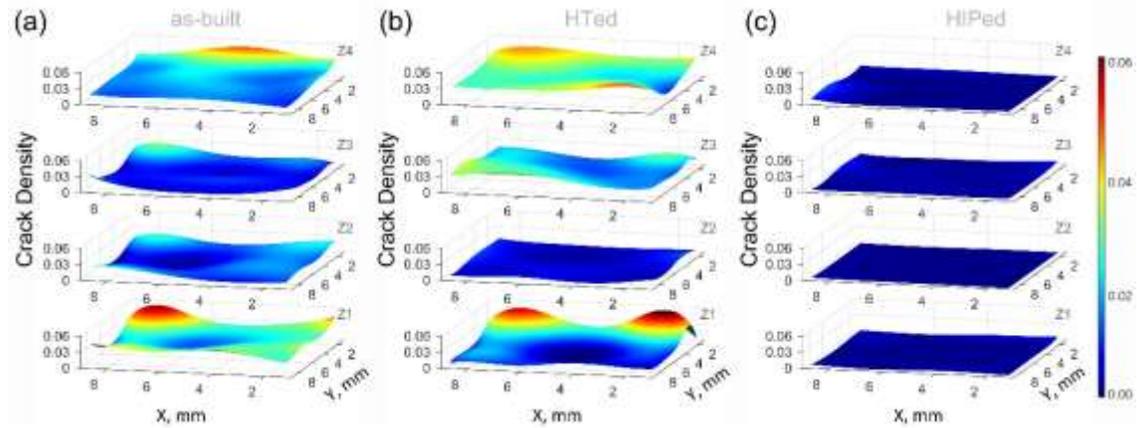

Fig. 8 Spatial distribution of crack density at different heights of sections (from Z1 to Z4) for (a) as-built, (b) anneal and (c) post hot-isostatic pressed specimens.

The dimensionless crack density for each image is calculated according to Eqn. (1) and (2), then arithmatically averaged for images recorded at the same height to get the mean values and errors. Fig. 9 shows the mean of the density of micro-cracks at different heights along the build direction . The automated grid and direct counting methods give similar values at all heights for the different specimens, confirming the accuracy and reliability of the automated method. It is worth noting that the larger error bars of mean crack density for both the as-built and annealed specimens (Fig. 9a) due to more drastic variation of cracks in these two conditions compared to the HIPed condition (Fig. 9a, b). Fig. 9a also shows the variation of crack density in the as-built and annealed were quite similar and followed the same trend with less cracks in the middle layers of builds.

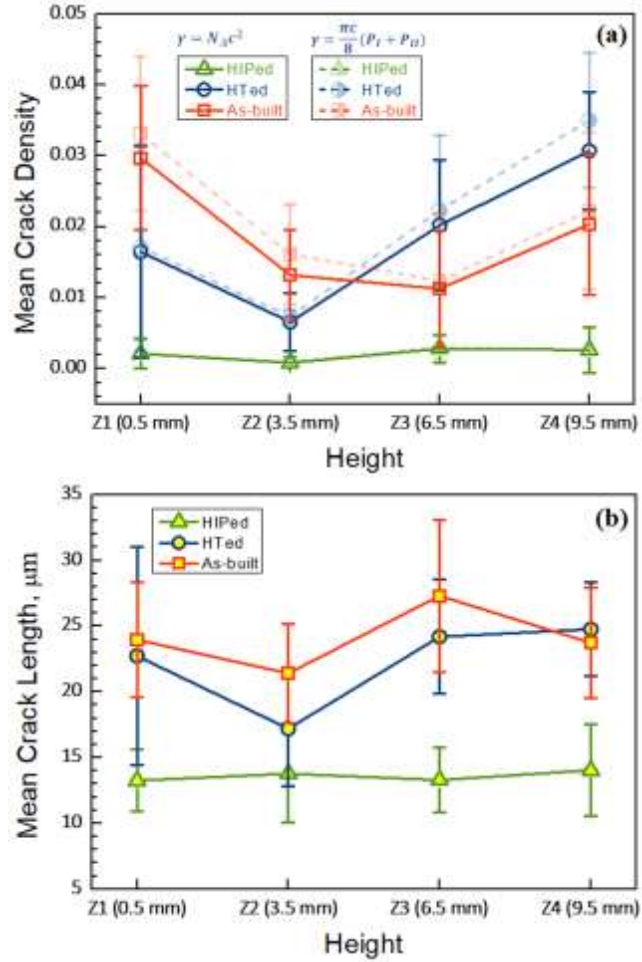

Fig. 9 Quantification of micro-cracks. (a) Variation of the mean value of micro-crack density and (b) mean crack length at different locations along the build direction for the three material conditions.

### *3.4. Calphad prediction of solidification behaviour*

Scheil-Gulliver solidification curve of the Hastelloy-X investigated in this study is shown in Fig. 10a in comparison with modified alloy by Harrison *et al.*, termed MHX [12]. The Hastelloy-X, investigated in this study, has a quite large solidification range of 270 K in comparison to 110 K of the MHX. Solidification gradient of the currently studied Hastelloy X is about 1054 K, while the MHX has a lower solidification gradient of 586 K. A lower absolute solidification gradient results in a higher resistance to solidification cracking [17].

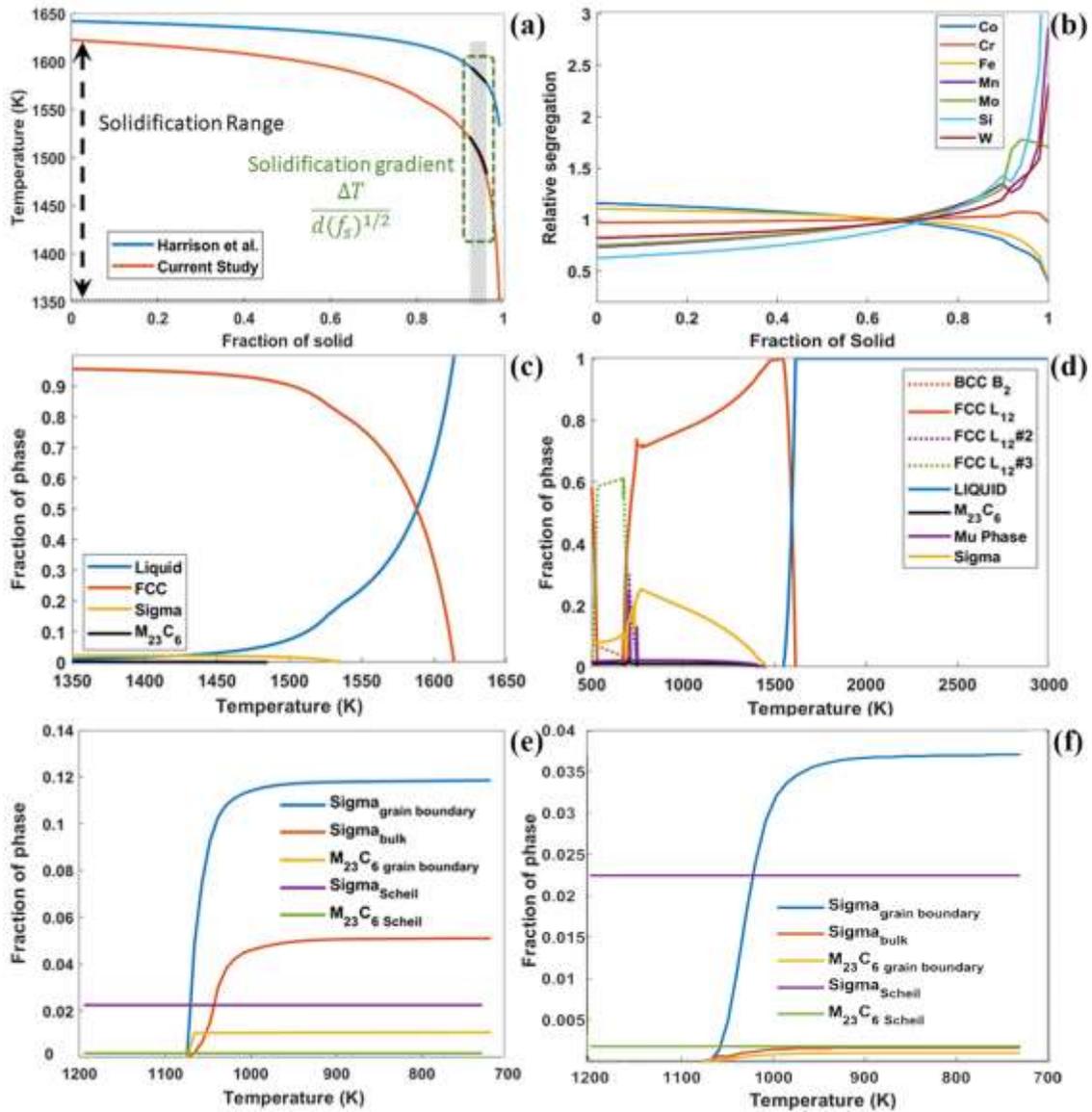

Fig. 10 Solidification behaviour of Hastelloy-X predicted by ThermoCalc. (a) Scheil solidification curve of the alloy investigated in this study and MHX by Harrison *et al.* [12] (b) Relative elemental segregation during Scheil solidification. (c) and (d) Phase fractions at different temperatures as expected during Scheil and equilibrium cooling respectively. (e), (f) ThermoCalc Prisma result for volume fraction of Sigma and $M_{23}C_6$ forming during cooling in HIP (0.2 K/s), (e) and annealing (8 K/s), (f), compared with Scheil solidification condition.

Fig. 10b shows the relative elemental segregation predicted by Scheil solidification, which can be interpreted as the compositional gradient in the solidification cell. The root (core) of a solidification cell forms first at the liquidus temperature at which the fraction of solid equals to 0, then the cell grows radially towards the cell boundary where the solid fraction is 1.0. Therefore, Fig. 10b shows the chemical segregation from the core of cells (corresponding to the fraction of solid 0.0) towards the cell boundaries (the solid fraction is 1.0). Elemental segregation directly affects the phase formation. As more segregating occurs towards the final stages of solidification, FCC structure cannot accommodate such enrichment in solutes and additional phases begin to form. In the case of Hastelloy X, small volume fractions of $\sigma$ (2.2%) and $M_{23}C_6$ (0.18%) phases were predicted to form at 1536 K and 1487 K, respectively in fast solidification (Fig. 10c).

Fig. 10d suggests that at 1450 K (temperature of the annealing) just the FCC phase is stable. Segregation from Fig. 10b was used as the starting point of the heat treatment simulation. Video S1 in the supplementary information shows the evolution of the elemental segregation for the duration of the heat treatment. Segregation was removed within a minute of the annealing. The second stage of the annealing, however, suggests that some $\sigma$ and $M_{23}C_6$ phases form upon cooling in the annealing and HIP (Fig. 10e, f). $M_{23}C_6$ formation was limited to grain boundaries, whereas $\sigma$ was predicted to form both on the grain boundaries and in the grain interiors. Under both cooling conditions in annealling and HIP, $\sigma$ and $M_{23}C_6$ started to form at about 1076 K. However, the equilibrium property diagram, Fig. 10d, predicts that $\sigma$ and $M_{23}C_6$ should be stable at 1442 K, suggesting that 366 K of undercooling was required to nucleate $\sigma$ and $M_{23}C_6$ phases. After passing 950 K, growth of all phases stoped in both cases. Cooling at a slower rate of 0.2 K/s in HIP allowed to reach a volume fraction of $\sigma$ and $M_{23}C_6$ of 0.17 and 0.01 respectively, approaching the thermodynamic maximum (Fig. 10d, e). A faster cooling rate of 8 K/s annealling reached only a volume fraction of 0.037 and 0.0016, respectively due to the limited kinetics (Fig. 10f). A much faster cooling rate is observed during LPBF process, $10^4 \sim 10^6 \, K \cdot s^{-1}$, resulting in even lower volume fraction of $\sigma$ and $M_{23}C_6$ forming (Fig. 10e, f).

*3.5. Microstructure characterization*

The microstructure in the as-built, annealed and HIPed conditions is shown in Fig. 11a – c, d – f and g – i, respectively, with increased magnification gradually. The as-built specimen shows a good consolidation with defects consisting of mainly micro-cracks marked by the white arrows and very few minute gas pores (Fig. 11a – c) in agreement with the observation in Figs 7 and 8. No clear evidence of precipitates was detected in the BSE imaging of the as-built sample (Fig. 11b, c) while the Scheil-Gulliver calculation predicted very small fractions of $\sigma$ and $M_{23}C_6$ (Fig. 10c). In contrast, the annealing heat treatment resulted in precipitation occurring in particular along grain boundaries (dash lines in Fig. 11f) and cracks (arrows in Fig. 11e) in consistent with Calphad simulation predicting the precipitation of $\sigma$ and negligible $M_{23}C_6$ (Section 3.4). Therefore, the observed precipitates were likely $\sigma$. HIPing resulted in quite similar microstructure to the annealed condition (Fig 11g – i). However, the density of precipitates in HIPed condition was higher than that in the annealed one (Fig. 11e, h) in agreement with Calphad prediction that HIP resulted in a higher volume fraction of $\sigma$ (Fig. 10e, f). Moreover, the precipitates formed both at grain boundaries and grain interiors in HIPed condition (Fig. 11i). Neither cracks nor lack-of-fusion (and keyhole) pores were detected in the HIPed condition apart from some minute round pores which were likely gas-entrapped (Fig. 11h).

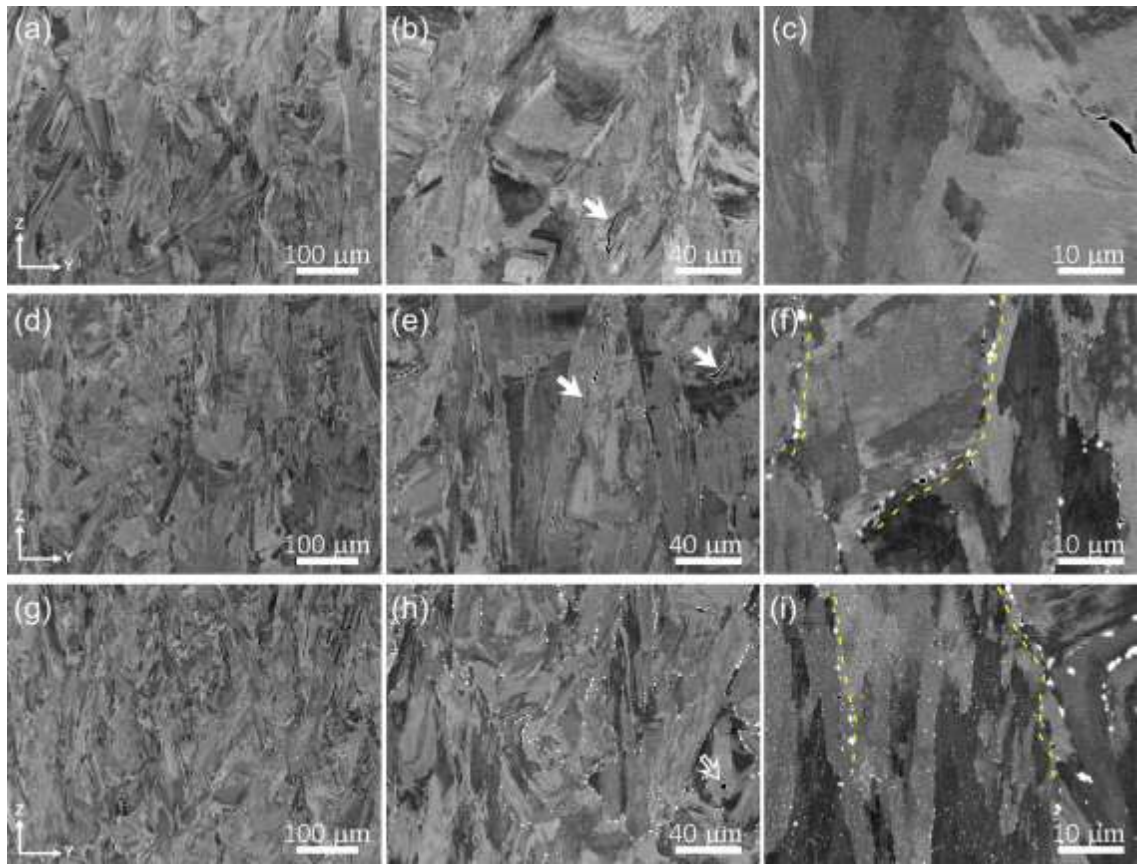

Fig. 11 SEM Back-scattered electron micrographs of (a – c) as-built, (d – f) post heat treated and (g – i) hot-isostatic pressed specimens. The solid arrows show the location of irregular cracks while the open one highlight a gas pore remained after HIP. Dash lines mark the grain boundaries.

The line EDS scanning of across crack in as-built specimen is shown in Fig. 12, from which the Si and O peaks and a slight C peak were detected, meaning higher Si, O and C concentration at the crack. Meanwhile, the Ni, Cr and Fe tend to be depleted at the crack edges.

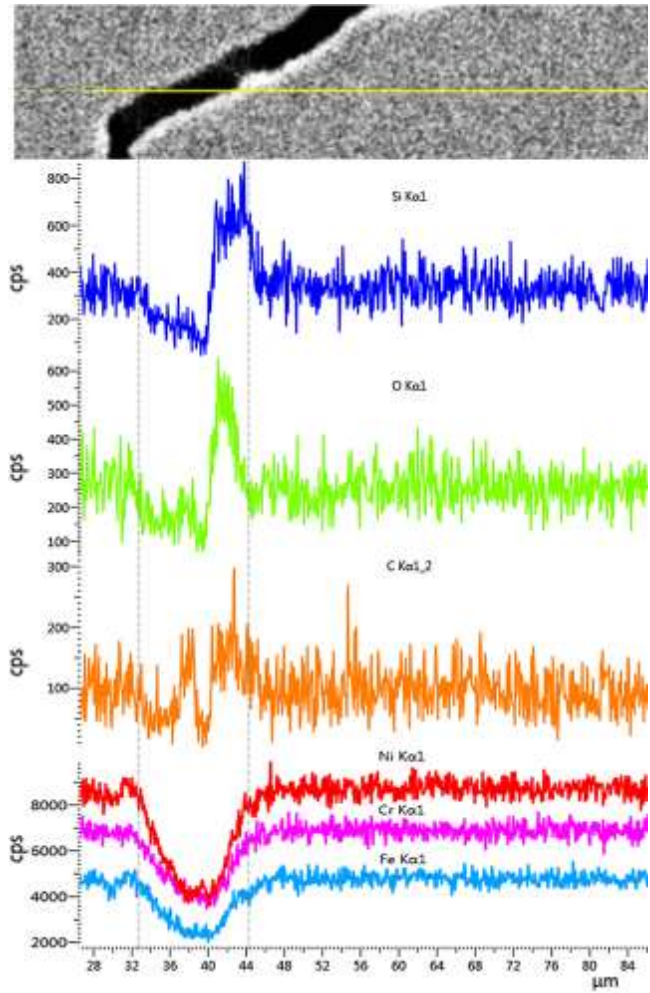

Fig. 12 EDS linear scanning across a crack in as-built specimen showing higher concentration of Si, O and C near the crack.

The grain microstructure including crystallographic orientation and grain morphology in the three different material condition are shown in Fig. 13 with a – c, d – f and g – i showing the inverse pole figure (IPF) map along the building direction, grain size and aspect ratio in the as-built, annealed and HIP conditions, respectively. The EBSD did not show strong preferred texture in all the material conditions. However, grains were quite elongated along the building direction (Fig. 13a, d, and g) consistent with previous studies [36, 37]. It is worth noting that the mean grain size in all the three conditions was quite similar with almost the same mean grain size (namely, 21.6, 20.1 and 22.0 µm for as-built, annealed and HIP samples, respectively) (Fig. 13b, e, and h). However, grains in the HIP condition tend to be more equiaxed than those in the other two conditions. The aspect ratio of grain was calculated to reflect the columnarity in grain morphology, which is defined as the ratio between the length of the major and minor axes of a grain. The average aspect ratio for as-built, annealed and HIPed samples are 3.5, 3.3 and 2.8 (Fig. 13c, f and i), suggesting recrystallisation might happen more significantly in HIP than in annealling, leading to more equiaxed grains in the HIPed condition.

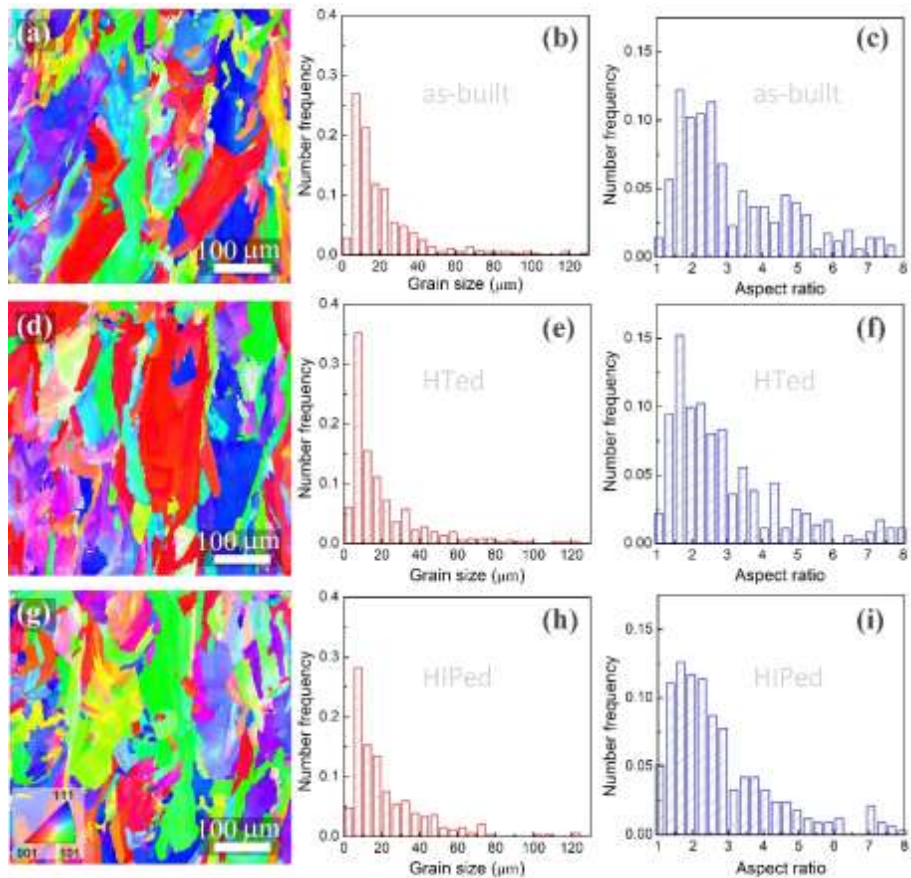

Fig. 13 EBSD IPF maps along buiding direction and the distribution of grain size and grain aspect ratio for (a – c) as-built, (d – f) annealed and (g – i) hot-isostatic pressed samples.

## 4. Discussion

### *4.1. Cracking mechanisms and distribution*

Hastelloy-X alloy is known to be susceptible to hot cracking (namely solidification cracking and ductility loss cracking) in LPBF [12]. While the solidification cracking are mainly influenced by solidification gradients and chemical segregation [14, 20], ductility loss cracking is governed by solid state phase transformation such as precipitation and to certain extent the chemical segregation [19]. The Scheil-Gulliver calculation of the solidification range and solidification gradient predict that the Hastelloy-X is susceptible to solidification cracking and more susceptible than the modified alloy studied by Harrison *et. al.* [12] (Figure 10a). Although the Scheil-Gulliver calculation predicts that there should be precipitation of $\sigma$ and $M_{23}C_6$ in agreement with a study done by Marchese *et al.* [19], SEM imaging did not find clear evidence of precipitates in as-built condition consistent with Harrison *et al.*'s findings [12]. However, the Scheil-Gulliver calculation usually over-estimate the chemical segregation due to an assumption that there is no diffusion in solids. The volume fractions of $\sigma$ and carbides in the as-built would be very small and negligible. SEM imaging of 2D sections might not be able to reveal the small fraction of precipitates. Because no significant precipitates were seen in the as-built condition (Fig. 11c), the solidification cracking due to the shrinkage and thermal contraction during solidification should be the main source causing the formation of cracks in the as-built alloy. Solidification cracking occurs at the terminal stage of solidification. Therefore, solidification cracks are often found at inter-dendritic regions and grain boundaries, explaining cracks in the as-built condition of the alloy were often seen at the boundaries between different cellular domains (Fig. 7). The Hastelloy-X has a steeper solidification gradient at the terminal stage of solidification than the modified alloy studied by Harrison *et al.* [12] (Fig. 10a). Therefore, the Hastelloy-X in this alloy should contain more solidification cracks. In addition, because the solubility in solids is lower than in liquid, there is chemical segregation from solid to liquid during solidification. As the interdendritic (or inter-cellular) regions were the last regions of solidification, there is enrichment of solute atoms at the boundaries of inter-dendritic regions. Fig. 10b shows that Si, Mn, W and Mo were most segregated in agreement with Marchese *et al.* [19] in which the Mo segregation was evidenced. Segregated regions (either enrichment or depletion) can weaken the alloy locally, making it more susceptible to cracking due to thermal stress in subsequently rapid cooling. Higher concentrations of Si and C towards the free surfaces of cracks found in the as-built condition (Fig. 12) indicate a significant role of the chemical segregation in formation of cracks.

The crack density distribution appears to be higher in the bottom, then decrease towards the middle, finally increase again in the top region of the build. As mentioned above (Section 3.1), higher cooling rates in the bottom part of a solid build than that towards the top of the build. Higher cooling rates cause faster and larger thermal contractions, inducing higher thermal stress that promote the formation of cracks in solidification. This probably explains the higher density of cracks observed in the bottom region in the as-built condition (Figs 8 and 9). In addition, tensile internal stresses tend to build up in regions towards the top and bottom layers while middle layers tend to be in compression [16, 38-40]. Compressive stresses in the middle of a build could help in closing the cracks, leading to a reduction of cracks in the middle, while tensile stresses increase the crack density towards the top of the build. More cracks were also seen near the free surface in the as-built condition (Fig. 8a). Fig. 5 shows that lower thermal conductivity (that should be the case near the free surface due to the surrounding powder) can lead to higher thermal gradient (Fig. 5d, h versus c, g). The higher thermal gradient could induce higher thermal stress, hence the crack density as seen in the as-built condition (Fig. 8a).

Post-processing treatments (such as annealing and HIP) are often used to enhance the mechanical properties via optimising microstructure and removing cracks. Such clustering of cracks observed in the as-built condition should remain unchanged during annealing without hot isostatic pressing as it was confirmed by Fig. 8b. The variation trend of crack distribution in the annealed condition is quite similar to that in the as-built (Figs 8a, b and 9). However, the combination of temperature and hydrostatic stress (*i.e.* HIP) can significantly close cracks and pores to achieve a near defect-free condition (Fig. 8c). This indicates that there was no crack re-healing during the annealing without pressing. This is probably because of oxidation of the free surface of defects, preventing the re-heal of cracks in annealing, making the closure of defects at high temperatures only possible if there was hydrostatic pressure applied.

*4.2. Influence of post-processing treatment in microstructure and mechanical properties*

Due to rapid cooling and high thermal gradient in additive manufacturing, materials fabricated by AM are reported to have different microstructure and higher mechanical strength than those made by other processes [36]. For example, Pham *et al.* observed AM 316L steel to have a strength which was about more than twice stronger than that of the conventional steel while the AM steel remains highly ductile thanks to a microstructure that promotes deformation twinning [41, 42]. High cooling rates and steep thermal gradient in powder bed fusion are believed to induce fine solidification cells of dense dislocations in AM alloys. Similar to cellular microstructure found in other AM alloys, Montero-Sistiaga *et al.* reported cellular structure of which the boundaries contain dense dislocations in Hastelloy-X alloy after printing [36]. They also showed that the microstructure was removed after heat treatment at a temperature of 1428 K. The heat treatment at a high temperature (1450 K) used in our study is in the range of temperature for recrystallisation which was also confirmed by Marchese *et al.*[43]. The recrystallisation often results in more equiaxed grains. This explains why smaller aspect ratios of grain morphology were seen in the heat-treated and HIP conditions (Fig. 13). The aspect ratio of grains in the annealed condition is still higher than that for the HIP condition, indicating a less degree of recrystallisation in annealing. Dislocations annihilated in recovery and recrystallisation in heat treatment at the high temperature of 1450 K, causing a decrease in the yield strength of the printed alloy after both the annealing and HIP (Fig. 6). Although precipitates were predicted by Calphad and observed via SEM for annealed and HIPed samples (Fig. 10e, f and Fig. 11f, i), the loss of the yield strength due to the disappearance of as-built dislocation microstructure should outweigh the hardening contribution of precipitates. Interestingly, the HIP specimen has higher yield stress as compared with the annealed one (Fig. 6). This is believed due to the precipitation strengthening as the precipitation occurred more dominantly in the HIP condition (Fig. 10e, h). Fig. 11f shows that more precipitation was mainly due to a slower cooling rate in HIP. According to the Orowan's hardening mechanism [44], precipitates can cause more significant hardening in the HIP material, leading to higher ultimate strength than the as-build and annealed condition as seen in Fig. 6a. HIP is also effective in closing pores and cracks, resulting in denser parts (Figs 8 and 9). The defects like micro-crack are detrimental to both the strength and ductility of metals [45]. In addition to the hardening induced by precipitates, the removal of cracks also contributes to the significant enhancement in strength (in particular the ultimate strength) in the HIP condition (Fig. 6a). The defect removal and higher hardening rate after HIP extended the uniform deformation before necking, hence increase the ductility of the HIPed alloy (Fig. 6a). The annealing without hydrostatic stress is not able to close cracks. Hence, there is no increase in the ductility observed for the annealed material.

5. Conclusions

This study identified optimized print parameters from literature data analysis. The identification was validated by fabricating Hastelloy-X with extremely high consolidation using the parameters. Most of process defects were identified to be caused by solidification cracking due to steep solidification gradient and solidification range as shown by Scheil-Gulliver calculation, and partly because of the chemical segregation. The distribution of crack density was quantified, showing higher densities in the bottom and top regions of builds. In addition, more cracks were detected near the free surface. Heat treatment without hydrostatic pressing (*i.e.* annealing) is not effective in removing the porosity, but results in significant decrease in yield strength due to the recovery and recrystallisation of microstructure at high temperatures. Although precipitation was predicted by Calphad and confirmed by SEM for the annealing, the yield strength was substantially reduced after annealing because the strength loss induced by the recrystallisation The unchanged density of cracks explains insignificant changes in the ductility of the annealed condition. By contrast, HIP – hot isostatic pressing (*i.e.* the use of heating and hydrostatic stress) is very effective in removing the process defects formed in the LPBF alloy to reach almost 100 % consolidation after HIP for the Hastelloy-X. The successful removal of defects and the more precipitation in HIP lead to significant improvement in uniform elongation (increasing from 13% to 20%), ultimate strength (from 965 MPa to 1045 MPa) while maintaining moderately high yield stress.

**Acknowledgment**

Much appreciated is the great support from Luke Griffiths for his original Python code package. L.C. acknowledge the support from National Key R&D Program of China (grant No. 2017YFB0305800). H.W. appreciated the support from the National Key R&D Program (Grant No. 2016YFB1100103) and National Natural Science Foundation of China (Grant No. 51771233). M.S.P thanks the support provided by the BIAM - Imperial Centre for Materials Characterisation, Processing and Modelling at Imperial College London.

[24] R.J. Williams, A. Piglione, T. Rønneberg, C. Jones, M.-S. Pham, C.M. Davies, P.A. Hooper, In situ thermography for laser powder bed fusion: Effects of layer temperature on porosity, microstructure and mechanical properties, Additive Manufacturing 30 (2019) 100880.

[25] E. Maire, P.J. Withers, Quantitative X-ray tomography, International Materials Reviews 59(1) (2014) 1-43.

[26] L. Griffiths, M.J. Heap, P. Baud, J. Schmittbuhl, Quantification of microcrack characteristics and implications for stiffness and strength of granite, International Journal of Rock Mechanics and Mining Sciences 100 (2017) 138-150.

[27] J.B. Walsh, The effect of cracks on the compressibility of rock, Journal of Geophysical Research (1896-1977) 70(2) (1965) 381-389.

[28] E.E. Underwood, Quantitative evaluation of sectioned material, Springer Berlin Heidelberg, Berlin, Heidelberg, 1967, pp. 49-60.

[29] H.S. Carslaw, J.C. Jaeger, Conduction of heat in solids, Oxford: Clarendon Press, 1959, 2nd ed. (1959).

[30] Principal Features of Hastelloy-X alloy, (accessed 01 August 2019). http://www.haynesintl.com/alloys/alloy-portfolio_/High-temperature-Alloys/HASTELLOY-X-alloy/HASTELLOY-X-principal-features.aspx.

[31] Product data of Hastelloy-X. http://www.spacematdb.com/spacemat/m-datasearch.php?name=Hastelloy%20X.

[32] M. Ohring, 9 - How engineering materials are strengthened and toughened, in: M. Ohring (Ed.), Engineering Materials Science, Academic Press, San Diego, 1995, pp. 431-500.

[33] H.-h. Ding, G.-a. He, X. Wang, F. Liu, L. Huang, L. Jiang, Effect of cooling rate on microstructure and tensile properties of powder metallurgy Ni-based superalloy, Transactions of Nonferrous Metals Society of China 28(3) (2018) 451-460.

[34] M.-S. Pham, B. Dovgyy, P.A. Hooper, C.M. Gourlay, A. Piglione, The role of side-branching in microstructure development in laser powder-bed fusion, Nature Communications 11(1) (2020) 749.

[35] P.A. Hooper, Melt pool temperature and cooling rates in laser powder bed fusion, Additive Manufacturing 22 (2018) 548-559.

[36] M.L. Montero-Sistiaga, S. Pourbabak, J. Van Humbeeck, D. Schryvers, K. Vanmeensel, Microstructure and mechanical properties of Hastelloy X produced by HP-SLM (high power selective laser melting), Materials & Design 165 (2019) 107598.
**33 / 34**

[37] O. Sanchez-Mata, X. Wang, J.A. Muñiz-Lerma, M. Attarian Shandiz, R. Gauvin, M. Brochu, Fabrication of crack-free nickel-based superalloy considered non-weldable during laser powder bed fusion, Materials 11(8) (2018) 1288.

[38] S. Zekovic, R. Dwivedi, R. Kovacevic, Thermo-structural finite element analysis of direct laser metal deposited thin-walled structures, Proceedings SFF Symposium,. Austin, TX, 2005.

[39] P. Mercelis, J.-P. Kruth, Residual stresses in selective laser sintering and selective laser melting, Rapid prototyping journal 12(5) (2006) 254-265.

[40] M.F. Zaeh, G. Branner, Investigations on residual stresses and deformations in selective laser melting, Production Engineering 4(1) (2010) 35-45.

[41] M.-S. Pham, P. Hooper, Roles of microstructures on deformation response of 316 stainless steel made by 3D printing, AIP Conference Proceedings, AIP Publishing, 2017, p. 040017.

[42] M.S. Pham, B. Dovgyy, P.A. Hooper, Twinning induced plasticity in austenitic stainless steel 316L made by additive manufacturing, Materials Science and Engineering: A 704 (2017) 102-111.

[43] G. Marchese, E. Bassini, A. Aversa, M. Lombardi, D. Ugues, P. Fino, S. Biamino, Microstructural evolution of post-processed Hastelloy-X alloy fabricated by laser powder bed fusion, Materials 12(3) (2019) 486.

[44] E. Orowan, Internal stress in metals and alloys, The Institute of Metals, London 451 (1948).

[45] L. Chen, L. Zhu, Y. Guan, B. Zhang, J. Li, Tougher TiAl alloy via integration of hot isostatic pressing and heat treatment, Materials Science and Engineering: A 688 (2017) 371-377.